\documentclass[12pt,preprint]{aastex}

\newcommand{\cx}{SN~2002cx}
\newcommand{\hk}{SN~2005hk}
\newcommand{\synow}{\textsc{Synow}}

\begin{document}

\shorttitle{SN 2005hk Spectropolarimetry}
\shortauthors{Chornock et al.}

\title{Spectropolarimetry of the Peculiar Type Ia SN 2005hk}

\author{Ryan Chornock\altaffilmark{1},
Alexei V. Filippenko\altaffilmark{1},
David Branch\altaffilmark{2},
Ryan J. Foley\altaffilmark{1},
Saurabh Jha\altaffilmark{1},
and Weidong Li\altaffilmark{1}}

\altaffiltext{1}{Department of Astronomy, University of California,
                 Berkeley, CA 94720-3411;
                 (chornock,alex,rfoley,saurabh,weidong)@astro.berkeley.edu}

\altaffiltext{2}{Department of Physics and Astronomy, University of Oklahoma,
  Norman, OK 73109; branch@nhn.ou.edu}

\begin{abstract}
We present Keck spectropolarimetry of the unusual type Ia supernova
(SN~Ia) 2005hk several days before maximum light.  An analysis of
the high signal-to-noise-ratio total-flux spectrum shows the object's
extreme similarity to the peculiar SN 2002cx.  SN 2005hk has an
optical spectrum dominated by \ion{Fe}{3} lines and only weak lines of
intermediate-mass elements, unlike a normal SN~Ia at this epoch.  The
photospheric velocity measured from 
the minima of strong absorption lines is very low for an SN~Ia
($\sim$6000 km s$^{-1}$), solidifying the connection to SN 2002cx.
The spectrum-synthesis code \synow\ was used to identify the
presence of iron-peak elements, intermediate-mass elements, and
possibly unburned carbon at similar velocities in the outer ejecta of
\hk.  Many weak spectral features remain unidentified.  The
spectropolarimetry shows a low level of continuum polarization
($\sim$0.4\%) after correction for the interstellar component and only
a weak \ion{Fe}{3} line feature is detected.  The level of continuum
polarization is normal for an SN~Ia, implying that the unusual
features of SN 2005hk cannot be readily explained by large
asymmetries. 
\end{abstract}

\section{Introduction}

Supernovae of type Ia (SNe~Ia) represent the thermonuclear explosion
of carbon-oxygen white dwarfs, probably as they approach the
Chandrasekhar mass in a binary system.  Despite the importance of
SNe~Ia as sites of iron-peak nucleosynthesis and their utility as
cosmological standard candles, the actual progenitor systems and
details of the explosion mechanism are still unknown; see
\citet{hn00} for a recent review of theoretical models and problems
associated with them. 

What is clear from an observational point of view is that most SNe~Ia
fall on a sequence between the peculiar SN 1991bg-like objects at one
end and the peculiar SN 1991T-like objects at the other \citep{bfn}.
The photometric sequence between these two endpoints \citep{ph93} is
paralleled by a spectroscopic sequence \citep{nu95}.  SN 1991bg
\citep{fi92b} was faint, red, had a rapidly declining light curve, and
showed prominent lines of \ion{Ti}{2} in its spectrum at maximum
light while SN 1991T \citep{fi92a,ph92} was bright, blue, had a
slowly declining light curve, and showed prominent \ion{Fe}{3} lines
in its spectrum.  \citet{nu95} explained the variation in normal
SNe~Ia between these two extrema as primarily a temperature effect,
driven by the amount of radioactive $^{56}$Ni formed in the explosion.
Most objects are normal, but estimates of the peculiarity rate vary
from 11\% to 36\% \citep{bfn,li01a}, depending on the definition of 
``peculiar.''

In recent years, some variants from this simple prescription for SN~Ia
diversity have been found \citep{li01b,li03,gar05}, including the
intriguing object \cx. 
\citet{li03} showed that \cx\ had \ion{Fe}{3} lines in its
pre-maximum-light spectrum like SN 1991T, a low total luminosity
comparable to SN 1991bg, the lowest photospheric velocities ever
measured in an SN~Ia at maximum light, an almost normal $B$-band light
curve but a bizarre plateau-like light curve in $R$ and $I$, peculiar
color evolution, and unusually rapid spectral evolution.  They
concluded that no existing theoretical model could match the
combination of features that \cx\ exhibited.  \citet{jha06}
showed that the peculiarities persisted to very late times.
The spectrum of \cx\ nearly 300 days after explosion was dominated by
a forest of narrow permitted \ion{Fe}{2} lines, each with a velocity
width of $\sim$700 km s$^{-1}$.  They also reported the detection of
low-velocity intermediate-mass elements and possibly \ion{O}{1}, with
important consequences for SN~Ia 
explosion models.  In addition, \citet{jha06} presented spectra
of several similar objects, including \hk, showing that \cx\ was not
unique and therefore a new subclass of SNe~Ia had been
overlooked until recently.

In this paper, we present Keck spectropolarimetric data of
\hk, a peculiar SN~Ia very similar to \cx, and an analysis of the
total-flux spectrum.  Full light curves and a spectral sequence will
be presented elsewhere (Phillips et al. 2006, in prep.).

\section{Observations}

SN 2005hk in UGC 272 was discovered (Burket \& Li 2005) at an 
unfiltered magnitude of $\sim$17.5 on 2005 October 30.25 (UT dates
are used throughout this paper) by the Lick Observatory Supernova 
Search with the 0.76-m Katzman Automatic Imaging Telescope (Li et al.
2000; Filippenko et al. 2001; Filippenko 2005).  An
independent discovery on 2005 October 28 was reported by the Sloan
Digital Sky Survey II team, who measured a \textit{g} magnitude of
18.8, two days after a non-detection (no upper limit was reported;
Barentine et al. 2005).  The host galaxy was observed as part of the
Sloan Digital Sky Survey (SDSS J002749.73-011200.0; Adelman-McCarthy
et al. 2006) and a redshift of $z = 0.0131 \pm 0.0002$ was 
measured.  We have adopted this redshift and removed it in all plots
in this paper. 

\citet{ser05} obtained an optical spectrum of \hk\ at Lick Observatory on
November 2, which showed the object to be an SN~Ia with
prominent absorption lines of Fe~III, similar to the peculiar object
SN 1991T.  The low photospheric velocities and faint
apparent magnitude led \citet{jha06} to classify \hk\ as a member of
an emerging subclass of \cx-like SNe~Ia.

We observed \hk\ on 2005 November 5.4 using the Low Resolution Imaging
Spectrometer (LRIS; Oke et al. 1995) on the Keck I 10-m telescope in
spectropolarimetry mode (LRISp)\footnote{See
\url{http://www2.keck.hawaii.edu/inst/lris/manuals/polarization/pol\_v3.ps}
for the online polarimeter manual (Cohen 1996).}.
We obtained a total exposure time of 5677~s in the four waveplate
positions ($\theta=0\degr,\ 45\degr,\ 22.5\degr,\ 67.5\degr$).  The
unusual total exposure time  
was due to poor guiding during the first exposure of the set, which
required us to abort and restart the waveplate rotation sequence.
After the end of the sequence, a short (200~s) exposure in the second
waveplate position ($\theta=45\degr$) was taken to complement the
aborted first exposure.  These two shortened exposures did not
substantially change the polarimetry if they were included in the
reductions, so we have used them in the analysis below.

Our instrumental setup employed both the blue and red 
sides of LRISp, with the D560 dichroic splitting the incoming light
beam.  The combination of the 600/5000 grism on the blue side and the
400/8500 grating on the red allowed us to cover a wavelength range of
$3170-9240$ \AA .  The 1.5\arcsec-wide slit gave a spectral
resolution of 6~\AA\ and 9~\AA\ on the blue and red halves of the spectrum,
respectively.  Conditions were good and the seeing was about
0.9\arcsec\ during the \hk\ observations.

Basic two-dimensional image processing was accomplished using
IRAF\footnote{IRAF is distributed by the National Optical Astronomy
  Observatories, which are operated by the Association of Universities 
  for Research in Astronomy, Inc., under cooperative agreement with
  the National Science Foundation.}, along with optimal extraction and
wavelength calibration of the one-dimensional spectra.  Our own IDL
procedures \citep{ma00} were used to flux-calibrate the spectra and
correct for telluric absorption.  The starting airmass was low (1.07),
but it climbed to 1.2 by the end of observations.  The slit was
oriented at a position angle (P.A.) of 130\degr\ to minimize
contamination from the host galaxy in background regions along the
slit. The parallactic angle over the course of the observations varied
from 10\degr\ to 50\degr, 
so differential light loss due to atmospheric dispersion \citep{fi82}
is visible in the individual spectra of the polarimetric sequence.  We
assumed the first complete observation had the correct spectral shape,
as it was taken at the lowest airmass and the P.A. was closest to
parallactic.  The spectral shape of the combined flux spectrum was
corrected to that of the first observation by means of a low-order
spline fit to the ratio of the two.

The spectropolarimetric reductions followed the basic procedure of
\citet{mi88}, as implemented by \citet{le01}.  The polarimetric
standard star HD 204827 \citep{sch92} was observed to determine the
P.A. offset of the instrument from the sky reference frame.  The blue
(red) side flux-weighted average polarization angle over the
wavelength range 3980$-$4920 \AA\ (5890$-$7270 \AA ) was set to the
catalogued value of 58.2\degr\ (59.1\degr ) for the $B$ ($R$)
band.  The derived offsets agreed to within 0.5\degr\ with
observations of the polarization standard HD 19820 \citep{sch92}.  Two
null standards, BD +32~3739 and HD 57702 \citep{sch92, mat70}, were also
observed and showed polarizations of $\lesssim$0.08\%.

\section{Results}

\subsection{Total-Flux Spectrum}

\begin{figure}
\plotone{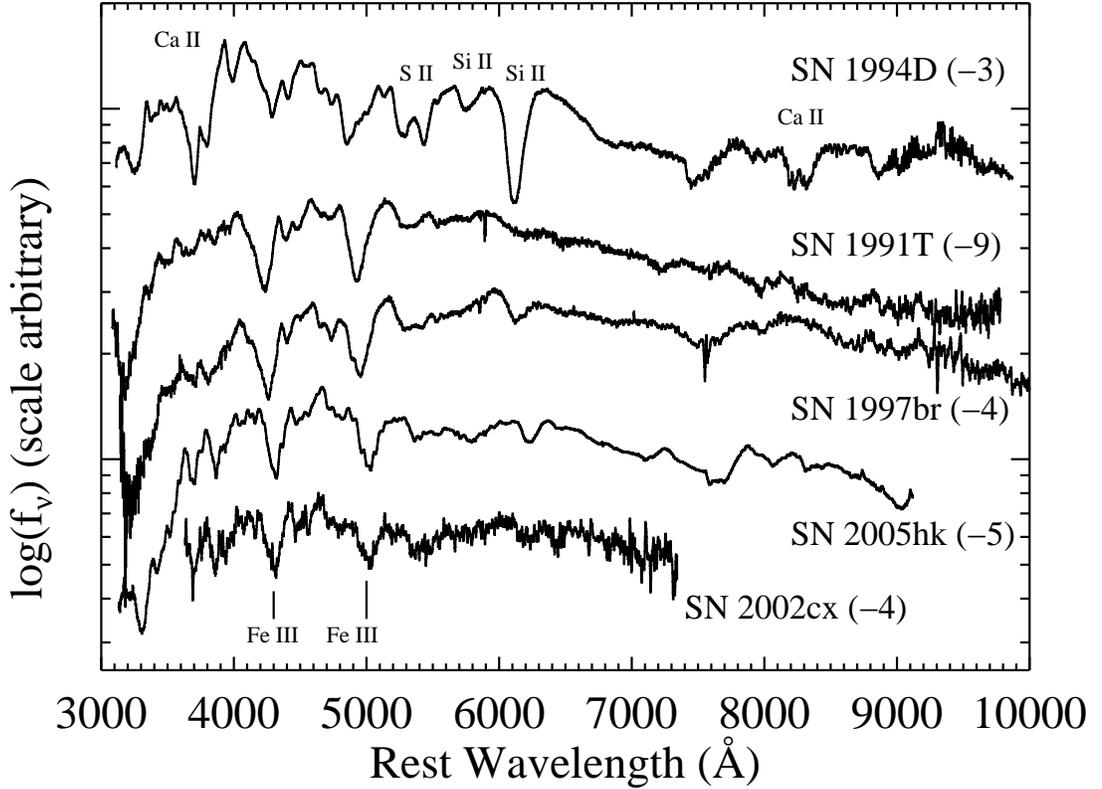}
\caption{Flux spectrum of SN 2005hk compared to the similar
  objects SN 2002cx (2002 May 17; Li et al. 2003), SN 1997br (1997
  April 16; Li et al. 1999), and SN 1991T (1991 April 19; Filippenko
  et al. 1992a).  Also plotted is a representative normal
  SN~Ia, SN~1994D (1994 March 17; Filippenko 1997).
  Narrow nebular emission lines from a superposed \ion{H}{2} region
  have been clipped from the SN 2002cx spectrum.  Numbers in
  parentheses are dates relative to $B$-band maximum. The strong
  \ion{Fe}{3} absorption lines are clearly more blueshifted in SN
  1991T and SN~1997br than in \hk.  Some strong absorption lines due to
  intermediate-mass elements (\ion{S}{2}, \ion{Si}{2}, and
  \ion{Ca}{2}) are marked on the SN~1994D spectrum, but they are weak or
  absent in the other four objects.
  For complete line identifications, see Figure~2.
}
\label{compfig}
\end{figure}

A by-product of the long exposure times necessary for
spectropolarimetry is the high signal-to-noise-ratio total-flux
spectrum.  In Figure~\ref{compfig}, the spectrum of \hk\ is
compared to early-time optical spectra of several other SNe~Ia.
The \cx\ spectrum is from four days before maximum light in the $B$
band \citep{li03}, about the same as this observation of \hk\ (W. Li
private com.; Phillips et al. 2006, in prep.), and appears very similar.

The most 
prominent features in the spectra of both objects are absorptions due
to \ion{Fe}{3} $\lambda\lambda$4404, 5129.  These \ion{Fe}{3} lines
are one of the distinguishing characteristics of the peculiar
SN~1991T-like class of objects, which are represented in
Figure~\ref{compfig} by both the prototype, SN 1991T \citep{fi92a}, and
the well-observed SN~1997br \citep{li99}.  One distinction
is that the \ion{Fe}{3} absorption minima in \cx\ and \hk\ are
observed to fall near 4320~\AA\ and 5030~\AA, implying unusually low
expansion velocities ($\sim$6000 km s$^{-1}$) for these SNe~Ia near 
maximum light \citep{li03,br04}.  

Normal SNe~Ia at this epoch also have
prominent spectral features due to intermediate-mass elements
(\ion{Si}{2}, \ion{Ca}{2}, \ion{S}{2}, etc.), as marked in
Figure~\ref{compfig} next to the spectrum of SN~1994D (Filippenko
1997).  These features are much weaker in the other objects, but are
clearly stronger in \hk\ than in SN~1991T.  They are also stronger in
this spectrum of \hk\ than in the Lick spectrum from three days
earlier presented by \citet{jha06}, indicating rapid spectral
evolution, as was observed in \cx\ at early times \citep{li03}. 

We used the parameterized supernova spectrum-synthesis code \synow\ 
\citep{fi97,fi99} to investigate the \hk\ spectrum more
quantitatively.  \synow\ uses several simplifying assumptions for  
computational ease, such as spherical symmetry, a blackbody-emitting
photosphere, and a resonance-scattering line source 
function in the Sobolev approximation.  Each ionic species is assumed
to have an optical depth that declines exponentially with velocity
above some photospheric velocity, $v_{ph}$, with an e-folding scale
$v_e$, and level populations in local thermodynamic equilibrium
(LTE) at some excitation temperature ($T_{exc}$).  The strength of
\synow\ is that of a fast line-identification tool, rather than ejecta
abundance determinations or detailed spectral modeling of the full SN
spectrum. 

\begin{figure}
\epsscale{0.85}
  \plotone{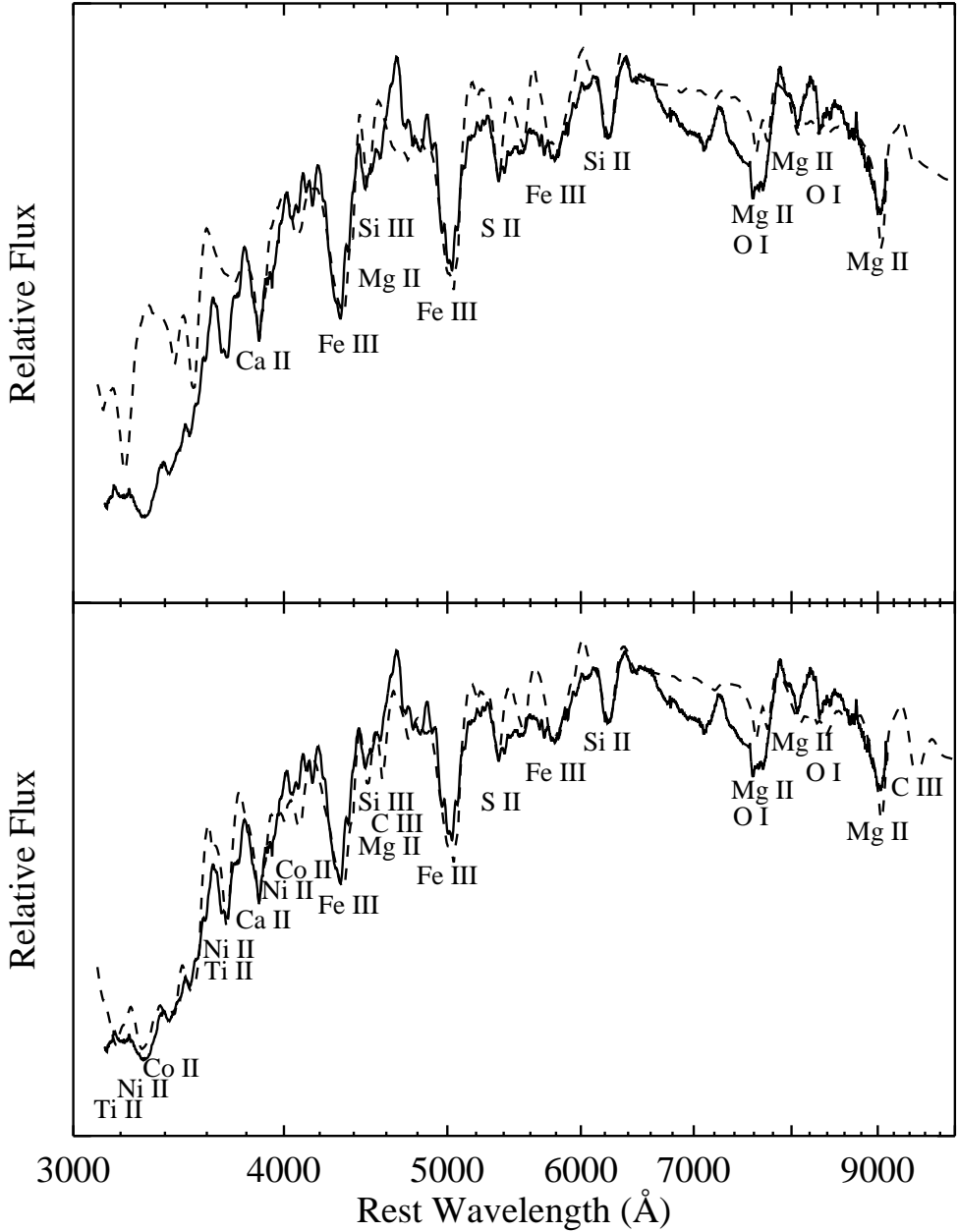}
\caption{\synow\ fits (dashed lines) to the flux spectrum of \hk\
  (solid line).  The fit presented in the top panel has seven
  ions we believe are present with high confidence.  The fit shown in
  the bottom panel is the same as that in the top, but with four
  additional species whose presence is less certain (\ion{Co}{2},
  \ion{Ni}{2}, \ion{Ti}{2}, and \ion{C}{3}).  The flux scale is linear
  and approximates $f_{\nu}$ 
  with an arbitrary tilt applied to better show spectral features and
  the wavelength scale is logarithmic to better show relative velocity
  widths of spectral features.  The primary species responsible for
  spectral features are marked.  See the text and Table~1 for \synow\
  fit parameters.}  
\label{synowfig}
\end{figure}

In the top panel of Figure~\ref{synowfig}, we present a fit using
seven ions that we 
believe are present in \hk\ with high confidence.  The photospheric
velocity used in the fit was 5000 km s$^{-1}$ and the ions were
assumed to have $v_e = 2000$ km s$^{-1}$, with no maximum velocity
imposed. (However, \ion{O}{1} and \ion{Mg}{2} are mildly detached,
$v_{ph} = 6000$~km~s$^{-1}$.) The low-ionization species (\ion{Ca}{2},
\ion{Mg}{2}, 
\ion{S}{2}, \ion{Si}{2}, and \ion{O}{1}) were fit using an excitation
temperature of 10,000~K, while a higher $T_{exc}$ of 15,000~K was used
for the doubly ionized species (\ion{Fe}{3} and 
\ion{Si}{3}).  The higher value of $T_{exc}$ was necessary to match
the strength of the \ion{Fe}{3} feature near 5900~\AA.  This synthetic
spectrum fails to fit some of the features of \hk, particularly at
bluer wavelengths, so we generated a second fit with four additional
ions (\ion{Ti}{2}, \ion{Ni}{2}, \ion{Co}{2}, and \ion{C}{3}), whose
presence we regard as possible, but not definite.  This second
synthetic spectrum is shown in the bottom panel of
Figure~\ref{synowfig}.  The additional  
ions are responsible for spectral features at desirable wavelengths,
as well as a decrease in the amount of near-ultraviolet (UV) flux, as 
discussed below.  A list of reference lines used in the fit and their 
optical depth can be found in Table 1.

\citet{br04} presented a \synow\ analysis of the \cx\ spectra.  Of
most relevance to us as a basis for comparison is the earliest
spectrum they analyzed, from one day before maximum light.  The best
fit for \cx\ has 
$v_{ph} = 7000$ km s$^{-1}$ and $v_e = 1000$ km s$^{-1}$.  While the
derived parameters imply a somewhat higher photospheric velocity in
\cx , $v_{ph}$ and $v_e$ are covariant and the two spectra show very
similar absorption minima, except that in direct comparison it is evident
that the minimum of the \ion{Si}{2} $\lambda$6355 line is at a lower
velocity in \hk .  The \cx\ \synow\ fit used six species
(\ion{Si}{2}, 
\ion{Si}{3}, \ion{S}{2}, \ion{Ca}{2}, \ion{Ti}{2}, and \ion{Fe}{3}), all
of which are present in our \hk\ fit.  The presence of \ion{Ti}{2} is
not certain in either fit, but \ion{Ti}{2} $\lambda$3760 matches the
absorption seen near 3680 \AA\ in both objects.  \citet{br04} observed
that this \cx\ fit could provide an adequate model for the SN
1991T-like object SN 1997br \citep{li99} if the photospheric velocity
were increased, indicating the kinship between these two classes of
supernovae. 

The \hk\ fit has five species not used for \cx .  Two of the ions,
\ion{O}{1} and \ion{Mg}{2}, have their strongest effects on the
spectrum between 7500~\AA\ and 9500~\AA , so they could not have been
detected at early times in \cx\ because the available spectra do not
extend as far to the infrared (IR).  The \ion{Mg}{2} $\lambda$4481 line
also makes a contribution in the blue, but it could have been
overlooked in \cx.  We also note that the \ion{Mg}{2} and \ion{O}{1}
lines in the red part of the spectrum are stronger than the
neighboring Ca IR triplet, which is opposite of most SNe~Ia
\citep{fil97}.  By comparison, the early spectra of SN 1991T are 
rather featureless at these wavelengths.  Oxygen and magnesium are
both products of incomplete carbon burning, so their presence could be a
diagnostic of the explosion process.  We cannot be certain that the
difference here between \hk\ and SN 1991T is solely due to differing
composition as opposed to greater velocity smearing or higher
temperatures and ionization levels, but more detailed study is
warranted.

In addition, \ion{Ni}{2} and \ion{Co}{2} are used in the second
\synow\ fit shown in Figure~\ref{synowfig}, but were not used in the
\cx\ fit. Their presence helps to fit features in the UV that were not
included in the spectral range of the early \cx\ spectra.  In
particular, the flux minimum  
near 3300~\AA\ is rather interesting.  SN 1991T showed a deep
absorption trough near 3200~\AA\ that was interpreted by \citet{fi92a}
to be the signature of \ion{Co}{2}, and hence of complete silicon
burning in the outer envelope of that object.  \citet{je92} showed
that while iron-peak elements are responsible for the UV opacity in
SNe~Ia, an atmosphere consisting solely of iron would have a flux
\textit{peak} here, while nickel and cobalt produce a minimum.
The more detailed models of \citet{maz95} confirmed and extended this
result.  By varying the composition of the atmosphere, they found that
deflagration products (taken from the popular W7 model of \citet{nom84})
were inadequate to fit the depth of this feature in SN 1991T at early
times and a substantial amount of $^{56}$Ni and its decay products was
necessary.  \synow\ is the wrong tool to use to determine abundances
in \hk, but the identification of \ion{Ni}{2} and \ion{Co}{2}
in this spectrum is suggestive of the presence of completely burned
material in the outer layers of \hk.

The presence or absence of carbon in the spectrum is particularly
interesting because of the implications for modeling the explosion of
\hk.  Some three-dimensional simulations of SN~Ia deflagrations have
shown that a significant amount ($\sim$40\%) of the carbon and oxygen
remains after explosion \citep{tra04}.  One potential explanation for
both the low luminosity and low expansion velocities of SNe 2002cx and
2005hk is that the explosions were very inefficient and an anomalously
large amount of fuel was left unburned.
At the conditions expected in an early-time SN~Ia atmosphere, 
\ion{C}{1} through \ion{C}{3} are the relevant ionization stages of
carbon \citep{ha99}.  Features attributed to  \ion{C}{2} $\lambda$6580 
and \ion{C}{2} $\lambda$7235 have been identified in the pre-maximum 
spectra of several normal SNe~Ia (e.g., SN 1998aq, Branch et al. 2003) 
and \ion{C}{1} has been identified in the near-IR spectra of the 
peculiar low-luminosity SN 1999by \citep{hof02}, though
\citet{mar06} placed strict upper limits on the presence of
\ion{C}{1} in the pre-maximum near-IR spectra of three
normal SNe~Ia.  Unfortunately, our spectra do not extend far enough to
the near-IR to test for the presence of \ion{C}{1} lines.

In the \synow\ fit presented in the bottom panel of
Figure~\ref{synowfig}, we have included \ion{C}{3} because 
an absorption near 4564 \AA\ is consistent with the strongest optical
line of \ion{C}{3}, the triplet at $\lambda$4647, blueshifted by
the photospheric velocity.  A similar feature was tentatively
identified as \ion{C}{3} in an early spectrum of SN~1999aa by
\citet{gar04}.  We regard this detection as possible because we cannot
claim an identification based on a single spectral feature to be
unambiguous, especially given the many as-yet  
unidentified lines.  The \synow\ fit shown in Figure~\ref{synowfig}
does show an additional \ion{C}{3} feature just outside of our
wavelength range at 9300 \AA, which might be observable in other
spectra or other \cx-like objects.  

\hk\ has a weak emission feature
at 7250 \AA\ with a corresponding P-Cygni absorption at 7100 \AA\ that
seems promising for \ion{C}{2} $\lambda$7235, but attempts to fit it
using \synow\ failed.  The \synow\ fits to $\lambda$7235 overpredict
absorption due to $\lambda$6580.  The observed spectrum does have an
absorption feature at the appropriate wavelength, but it is much too
weak to match the model.  LTE models generically predict that
$\lambda$6580 should be stronger than $\lambda$7235 because the lower
levels of $\lambda$7235 are the upper levels for
$\lambda$6580.  Likewise, the synthetic spectra with \ion{C}{2} predict
an absorption due to the $\lambda$4745 multiplet which falls on an
observed emission feature at 4662 \AA.  The two longer-wavelength
features are either not due to \ion{C}{2} or they are far out of LTE,
but in any case we cannot identify \ion{C}{2} with certainty.
\citet{je92} encountered similar difficulties with the 
identification of \ion{C}{2} in the early spectra of the more normal
SN~Ia 1990N (but see Mazzali 2001).  The wavelength coincidence of
features in the observed spectrum with the strongest expected lines of
\ion{C}{2} and \ion{C}{3} is suggestive, but we cannot claim a firm
detection of any form of carbon at this time. 

\begin{figure}
\plotone{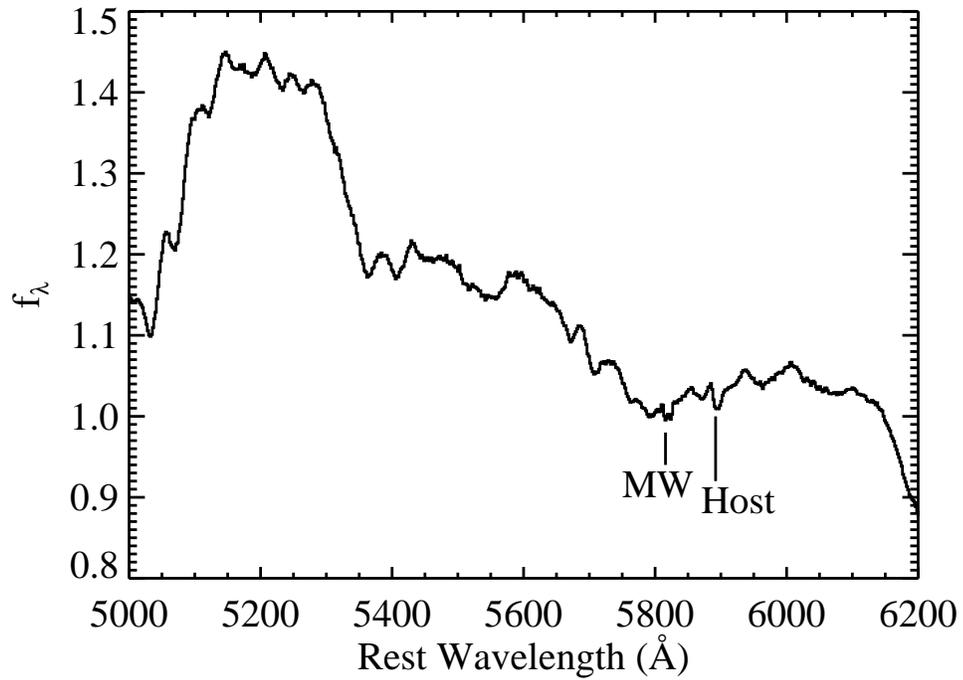}
\caption{Section of the SN 2005hk spectrum enlarged to
 better show the numerous unidentified weak spectral features.  The
 weak interstellar Na~I~D lines from our Galaxy (MW) and the host
 galaxy of SN~2005hk are labeled.} 
\label{nafig}
\end{figure}

One last unusual aspect of the \hk\ spectrum is the presence of many
low-amplitude, high-frequency spectral features, particularly at
wavelengths less than $\sim$6000~\AA.  See Figure~\ref{nafig} for a
closer look at some of these features, which have amplitudes of a few
percent of the continuum over scales of 20--30~\AA.  Despite the eleven
species present in the synthetic spectrum, it does not reproduce this
small-scale structure. \citet{br04} noted the existence of many weak
features in the pre-maximum spectra of \cx.  They were also unable to
fit these features with \synow\ and speculated that either non-LTE
level populations enhanced otherwise weak lines or
clumpy velocity structure in the ejecta might be responsible.  The 
\hk\ spectrum shows many of the same features, plus others visible
due to its higher signal-to-noise ratio, making clumps at 
\cx-specific velocities seem less likely.  In addition, if the ejecta
were strongly clumped, we might expect to see some polarimetric
signature, as in SN~2004dt (Wang et al. 2006; see below).  We do not
at this time
have solid identifications for these features, but their confinement
to wavelengths blueward of 6000 \AA\ suggests an origin in iron-peak
elements.  It is conceivable that 
many of them are also present in other SNe~Ia, but are
blended together in objects with higher expansion velocities.

\subsection{Spectropolarimetry}

\subsubsection{Correction for ISP}

Before analyzing the polarimetry, we must correct the observations for
the effects of interstellar polarization (ISP), both in our own Galaxy
and in UGC 272.  The expected Galactic 
contribution to the ISP is low, as the Galactic latitude is high
($-63.5$ $\deg$) and the dust maps of \citet{sfd98} predict a Galactic
reddening, $E(B-V)$, of only 0.023 mag.  The polarization efficiency
($p_{MAX}$ / $E(\bv\ )$) of Galactic dust has been found to be
typically 3\% mag$^{-1}$, with a maximum value of 9\% mag$^{-1}$
(Serkowski et al. 1975).  We observed two stars, HD 2158 and
HD 2636, to act as probes of the actual Galactic ISP. 
Following the prescription of \citet{tr95}, both stars are within
0.6\degr\ of \hk\ and have spectroscopic parallaxes that place them 
more than 150~pc above the Galactic plane and thus sample almost
all of the Galactic dust column along that line of sight.  According
to the discussion above, we expect a polarization of about 0.07\%,
with a maximum of 0.2\%.  The actual debiased, flux-weighted
$V$-band polarizations we measure for HD 2158 and HD 2636 are 0.14\%
at $\theta = 131$\degr\ $\pm$ 2\degr\ and 0.25\% at
$\theta = 130$\degr\ $\pm$ 1\degr , respectively.

Lacking a reason to prefer one star over the other, we averaged the two
together to form our estimate of the Galactic contribution to the ISP,
and smoothed versions of $q$ and $u$ were subtracted from the
observations of the supernova.  This non-parametric procedure was
followed because at this  low polarization level we must allow for
the potential effects of instrumental polarization which might cause
the polarization to not follow the standard wavelength dependence of
\citet{ser75}.  The null standards, as noted above, had measured
polarizations below 0.08\%, but we have seen instrumental polarization
artifacts at the 0.1\% level with LRISp in the past, particularly on the 
blue side of the spectrograph \citep{le02}.  The probe stars were
observed immediately after \hk\ and with the slit at the same position
angle, and thus should have similar amounts of instrumental
polarization, if any.  Any uncertainties in the removal of Galactic ISP
are folded into the estimation of the host-galaxy contribution to
ISP below.

\begin{figure}
\plotone{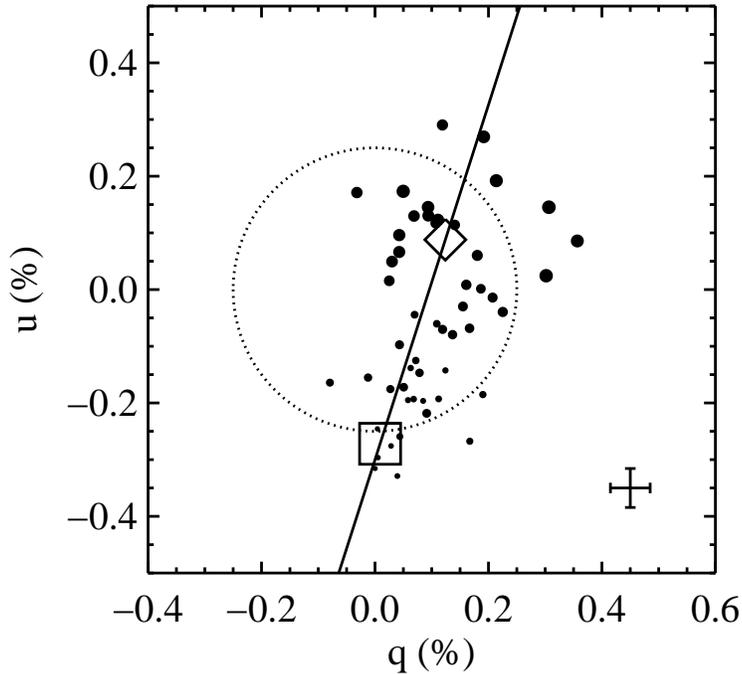}
\caption{Polarization data for SN~2005hk (after correction for Galactic
  ISP) are plotted in the $q-u$ plane. The data have been binned to
  100~\AA\ for clarity, and only points with observed wavelengths
  between 4000~\AA\ and 9000~\AA\ are shown here.  The symbol size is
  proportional to wavelength, so bins at longer wavelengths are
  represented by larger points.  Bluer-wavelength bins
  tend to be found in the cloud of points at lower values of $u$ than
  redder ones.  The dotted circle represents the typical magnitude of
  ISP expected from the host galaxy based on analysis of the Na~I~D lines.
  The square marks the location of our chosen ISP vector, and the
  diamond marks the location of our continuum measurement (7000--7500~\AA).  
  The line between them represents the
  direction of the new $q$ axis after subtracting the ISP vector and
  rotating the Stokes parameters about the new origin at the square.
  See text for details and Figure~\ref{spolfig} for the 
  spectropolarimetry after ISP subtraction and rotation.  The cross at
  (0.45,$-0.35$) 
  shows the median statistical error bars for the points in the plot.}
\label{qufig}
\end{figure}

It is instructive at this point to examine the location of the \hk\
spectropolarimetric data points in the $q-u$ plane, as shown in
Figure~\ref{qufig}.  The data have been binned up to 100 observed \AA\
per point for clarity, and noisy data at the ends of the spectrum are
not plotted in this figure.  The first important fact is that
the overall polarization is low, less than 0.5\%.  The second is that
the points do not lie randomly in the plane.  The distribution of
points is elongated in a direction which does not intersect or point to the
origin.  This clearly means that host ISP cannot be the sole source of
polarization in \hk\ \citep{ho01,le02}.  The effect of vector
subtraction of a small value of ISP is similar to changing the origin
of the $q-u$ plane, as the wavelength dependence of ISP is not strong 
\citep{ser75}.  Furthermore, the data points at shorter wavelengths
are systematically located below those at longer ones, indicating a
gradient in polarization percentage in this object.

The contribution of dust in the host galaxy of \hk, UGC 272, to the
total ISP is harder to determine than that of our Galaxy.  The
magnitude of the expected 
polarization can be estimated as above from the reddening, and the
reddening in turn can be estimated from the interstellar Na~I~D
$\lambda\lambda$5890, 5896 absorption lines.  Absorption due to Na~I~D
can be seen in Figure~\ref{nafig} at the redshift of UGC 272, with an
equivalent width (EW) of $\sim$0.35~\AA. \citet{bar90} 
found a crude relationship between the 
reddening and the total equivalent width of Na~I~D absorption,
$E(B-V)=0.25$~mag~\AA$^{-1}$ $\times$ EW(Na~I~D).  This, combined with the
\citet{ser75} relationship for polarization efficiency, yields an
estimate for the typical expected host ISP of 0.26\%.  

\citet{munz} 
have also derived a relationship for the reddening as a function of
the EW of the D1 component of Na~I absorption, but the two components
of the host Na~I~D absorption are not resolved in our spectrum.  The
sightlines in the \citet{munz} 
study showed a variation in the ratio of EWs for Na~I D1:D2 of 1.1:1
to 2:1.  Using these values as a bracket for the true value, the range
of expected polarizations is 0.18--0.24\% for the \citet{munz}
relationship, consistent with that predicted using the \citet{bar90}
relationship.  The dotted circle shown in Figure~\ref{qufig} has a
radius of 0.26\% to show the magnitude of the expected host ISP
relative to the data, though dust with maximal polarization efficiency
could show three times as much polarization.

The Na I D methods described above can only estimate the magnitude of
host ISP, but not the direction.  Fortunately, we can use the \hk\
polarimetry itself to estimate the host ISP.  The optical spectrum of
an SN~Ia at blue wavelengths is expected to have almost no
continuum polarization on theoretical grounds.  Opacity due to many
overlapping spectral features from iron-peak elements should act to
depolarize the continuum, even if the continuum is intrinsically
polarized by electron scattering in an aspherical atmosphere
\citep{ho01}.  Different authors in the 
literature have chosen different methods to take advantage of this
assumption. \citet{ho01} used an ISP that made the polarization at the
(arbitrary) blue end of their spectra almost zero.  \citet{wa06} and
\citet{leo05} used the polarization at specific wavelengths of blue
($\lambda<$~5000 \AA) emission features to define an ISP under the
assumption that the observed polarization of these emission features
is entirely due to ISP.  We adopt a spectral window of 4000--4200~\AA\ 
in the SN rest frame to measure our host
ISP.  The exact endpoints of this window are arbitrary, but this
region was chosen to avoid the strong \ion{Ca}{2} and \ion{Fe}{3}
absorption lines at nearby wavelengths which could potentially show
polarization features.  Our measured value for ($q_{ISP}$,$u_{ISP}$)
of (0.009\%, $-0.272$\%) has a formal uncertainty of 0.018\% in each
Stokes parameter.  This point is marked by a square
in Figure~\ref{qufig}, and is consistent with our expectations from the
reddening arguments above.  We have adopted a \citet{ser75} law with
$R_V = 3.1$ to subtract this ISP on the assumption that dust in
UGC~272 resembles Galactic dust.

As an aside, we note that it is possible to use these polarization
measurements as a constraint on the reddening to \hk.  One possible
explanation for the low apparent luminosity of SN 2002cx-like objects
is extinction, though this is hard to reconcile with the relatively
blue early-time \bv\ color of SN~2002cx \citep{li03}.  We have adopted a
host ISP of 0.272\%, which for standard dust polarization efficiencies
corresponds to $E(\bv)$=0.091~mag.  If this is combined with the
Galactic contribution and $R_V$ of 3.1 is adopted, then we estimate a
total extinction of $A_V \approx 0.35$ mag, not nearly enough to conclude
\hk\ was an SN~Ia of intrinsically normal luminosity (Phillips et
al. 2006, in prep.).

\subsubsection{SN 2005hk Polarization}

The continuum polarization of SNe~Ia can be understood as the result
of electron scattering in an aspherical supernova atmosphere
\citep{ho91,je91}.  If the projection of the supernova on the sky
were circularly symmetric, the local polarization vectors would cancel
each other 
when averaged over the unresolved object and no polarization would be
measured.  Therefore, the continuum polarization level represents a
meaningful measurement of the asphericity of the supernova.  A number
of authors have used the spectral region near 7000 \AA\ to define a
continuum polarization because of the relative lack of strong lines
\citep{ho01,ka04,leo05}.  
Here we choose the (rest-frame) wavelength range 7000--7500~\AA\ to
define a continuum region in \hk.  The wavelengths were chosen to
avoid the \ion{O}{1} $\lambda$7774 absorption trough.  We measure
Stokes parameters of (0.124\%, 0.088\%) for this region prior to ISP
subtraction, and mark this point with a diamond in Figure~\ref{qufig}.
After subtraction of our chosen ISP, this continuum region has a
polarization angle of 35.6\degr\ at a level of 0.36\%, an entirely
typical value for a normal SN~Ia, as discussed below.

The one-dimensional presentation of spectropolarimetric data is
problematic because of the inherent two-dimensional nature of the
measured Stokes parameters.  In addition, at low polarizations the
formal polarization ($P = (q^2 + u^2)^{\frac{1}{2}}$) is
statistically biased.  One solution is to rotate the $q-u$ coordinate
system by a convenient amount to new, rotated Stokes parameters (RSP)
chosen to align one Stokes parameter (called $q_{RSP}$ here) with the
axis of the system.  In this new coordinate system, $q_{RSP}$ is an
estimator of the total amount of polarization aligned with the axis of 
symmetry and $u_{RSP}$ can show deviations from a single axis of
symmetry in the system.

\begin{figure}
\epsscale{0.65}
\plotone{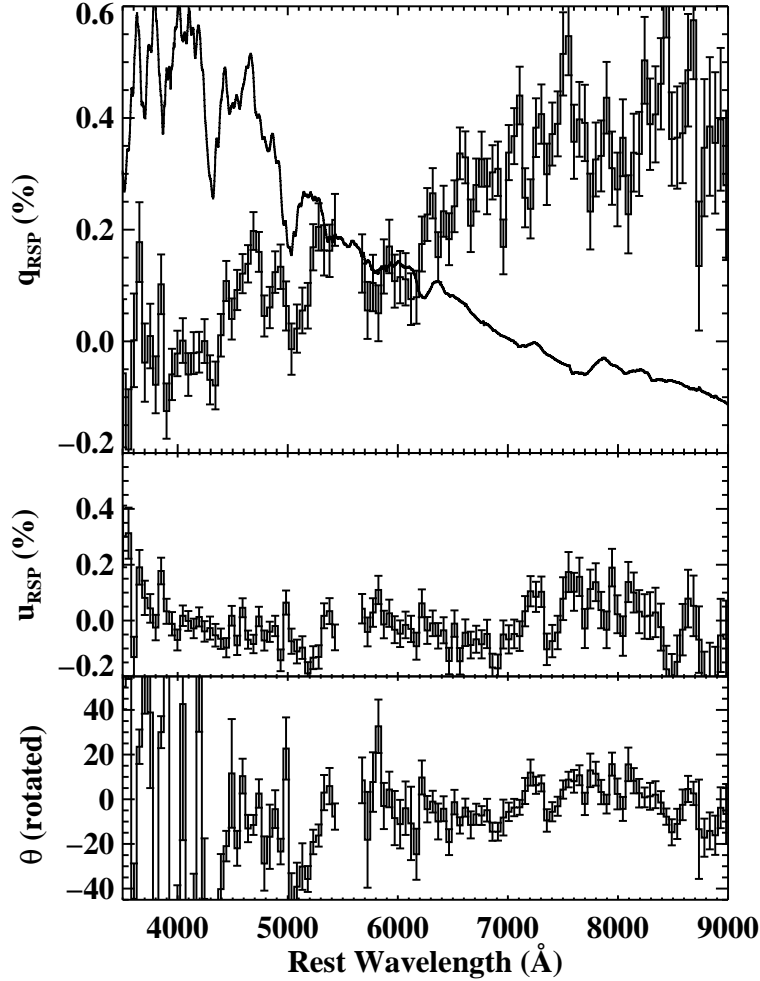}
\caption{Polarization data for SN 2005hk after removal of ISP and
  rotation about the new origin to align the rotated Stokes parameter
  $q_{RSP}$ with the continuum axis of symmetry.  The flux spectrum 
  (scale is linear $f_{\lambda}$) is overplotted on the $q_{RSP}$
  data.  The 
  polarization data are presented in 50~\AA\ bins.  A continuum
  polarization of 0.4\% can be seen in addition to a feature at the
  Fe~III line near 5000~\AA.  Any polarization features in $u_{RSP}$ 
  would represent deviations from a single axis of symmetry.  The
  relatively flat $\theta$ curve over 
  those wavelengths where the polarization angle is numerically
  well-defined confirms that SN 2005hk has a single dominant axis of
  symmetry.  The small gap in the data near 5600 \AA\ is the region of
  the spectrum affected by the dichroic and where small spurious
  polarization features were present.
}
\label{spolfig}
\end{figure}

For \hk, we have chosen to rotate the Stokes parameters by 35.6\degr\
to align $q_{RSP}$ with our chosen red continuum window of 7000--7500~\AA.  
These new rotated Stokes parameters are presented in
Figure~\ref{spolfig} and show a continuum polarization rising from zero 
near 4000~\AA\ (by construction for this choice of ISP) to $\sim$0.4\%
at 9000~\AA.  The flatness of the $\theta$ curve and the lack of
significant features in $u_{RSP}$ show that a single
axis of symmetry dominates the continuum polarization.  Note that at
very low polarization levels, such as seen at wavelengths less than
5000 \AA\ in our data, the angle of polarization becomes ill-defined.
No strong line features are seen in the polarimetry, but there does
appear to be a weak polarization modulation at the \ion{Fe}{3}
$\lambda$5129 line.  A 0.2\% depression in $q_{RSP}$ indicates some
depolarization and a smaller feature in $u_{RSP}$ may indicate a
rotation of the axis of symmetry.  Other, smaller polarimetric
features may also be present, but we caution against
over-interpretation of weak ($\lesssim$0.1 \%) spectropolarimetric
features.

\section{Discussion}

Early attempts to measure broad-band polarization of SNe~Ia were
unsuccessful, both due to the low intrinsic polarization levels and
the difficulty of disentangling the effects of ISP, so intrinsic
polarization remained undetected until SN 1996X showed marginal
evidence at the $\sim$0.3\% level \citep{wwh97}.  Since then,
intrinsic polarization has been detected in a number of SNe~Ia, both in
the continuum and in line features \citep{ho01,ka03,wa03,leo05,wa06}.

There are no published spectropolarimetric data for SN 1991T-like or
\cx-like objects to act as a basis for comparison with \hk, but there
are a few relevant measurements.  SN~2004bv had an
early-time spectrum that showed prominent \ion{Fe}{3} lines like SN
1991T \citep{fo04}, and \citet{pm05} obtained a single epoch of
$R$-band imaging polarimetry two weeks after maximum light.  They
found a polarization of 0.11\% $\pm$ 0.02\%, consistent with the
polarization of the foreground stars in their field and therefore with
low intrinsic supernova polarization.  \citet{leo00} reported on
maximum-light spectropolarimetry of the unique SN~Ia 2000cx, which
resembled the SN 1991T-like class of objects \citep{li01b}.  They
found a continuum 
polarization of $\sim$0.5\%, with a 0.3\% modulation across the
\ion{Si}{2} line, but most of the continuum polarization may be due to
ISP (D. Leonard, private com.).  SNe~Ia with normal spectra and
luminosities have shown continuum 
polarizations at the level of $\sim$0.3\% \citep{wwh97, wa03, leo05}.
Two subluminous SNe~Ia, SN 1997dt and SN 1999by (which resembled SN
1991bg), have shown somewhat higher continuum polarizations, up to 0.5\%
and 0.8\%, respectively, with the exact values dependent on
assumptions about ISP and the limited wavelength range of the
observations for these two objects \citep{ho01,leo05}.

In summary, the magnitude of the continuum polarization in \hk\ is
entirely consistent with observations of other SNe~Ia.  Core-collapse
supernovae, by contrast, show polarizations of up to a few percent
(e.g., Leonard \& Filippenko 2005).  Current evidence seems to
indicate that the core-collapse supernova explosion mechanism is
fundamentally aspherical, and objects with lower polarization levels
are those with extensive hydrogen envelopes that act to damp and hide
asymmetries.  The low polarization of thermonuclear supernovae
indicates a more nearly spherical explosion mechanism.  Whatever
unknown mechanisms are generating asphericities in the explosions of
normal SNe~Ia, such as instabilities in the growth of a deflagration
front (e.g., Ghezzi et al. 2004), are probably also responsible in \hk.

The continuum polarization percentage in some other SNe~Ia varies with
wavelength.  The most striking example is the low-luminosity SN~1999by
\citep{ho01}, whose polarization rises dramatically to the red, as it
does to a lesser extent in \hk.  Thomson scattering is 
independent of wavelength, so the 
slope in polarization percentage is interpreted as a signature of the
depolarizing effect of line opacity, which is stronger at UV and blue
wavelengths than in the red \citep{ho01}.  This result is dependent on
the choice of ISP, which has been driven by the expectation of low
intrinsic polarization at blue wavelengths.  SN~2001el provides a note
of caution, as the ISP derived from the polarization of blue emission
features at early times \citep{ka03} does not agree with the
polarization at late times \citep{wa03} when the intrinsic
polarization should be low.  We do not have multiple epochs of 
polarimetric data on \hk, so we cannot use the late-time polarization
to test our choice of ISP.

A number of SNe~Ia have shown definite spectropolarimetric line
features, but in \hk\ they are weak.  The strongest line
features ($P\lesssim$2.5\%) have been seen in those SNe~Ia that show
high-velocity absorption at early times in the Ca IR triplet and the
\ion{Si}{2} lines.  \citet{ka03} showed that partial obscuration of
the photosphere by clumps of intermediate-mass elements at high
velocity provides an adequate description of 
the polarimetric data for SN~2001el, perhaps due to a gravitationally
confined detonation \citep{kp04}.  \citet{wa06} preferred to explain
the strong polarization features of SN~2004dt by clumpy protrusions of
silicon into an oxygen-rich layer in the outer SN atmosphere, maybe
as a result of turbulent 
burning processes during the explosion.  \hk\ does not show
high-velocity absorptions and the \ion{Si}{2} and \ion{Ca}{2} lines
are weak, so these strong polarization modulations are not
expected.  We can at least conclude that the lack of strong
polarization increases in the absorption troughs of \hk\ limits the
presence of inhomogeneous clumps above the photosphere.

SNe~Ia lacking the high-velocity material show much smaller line
polarization features, usually at the same level as the continuum
polarization or less \citep{ho01,leo05}.  \citet{ho01} showed that, in
general, strong absorption lines in SNe~Ia should be associated with
polarization minima as most photons have been absorbed and reemitted
in the line, destroying the directional information imprinted by
electron scattering.  The spectral
features in \hk\ may be just too weak to leave strong polarization
signatures.  For example, some of the strongest polarization features
have been seen in other SNe~Ia in the \ion{Si}{2} $\lambda$6355
absorption line, but 
in \hk\ the absorption depth is only $\sim$12\% of the continuum.
In the simple aspherical toy models of \citet{lf01}, such a feature
would only be 
expected to result in a polarization feature that was at most
$\sim$12\% of the continuum polarization level.  This model
does not take into account optical depth effects, clumping, or
multiple overlapping lines, but it may allow us to understand the lack
of strong spectropolarimetric line features in \hk.  For 
a continuum polarization of 0.4\%, this model would predict a
polarization modulation in the \ion{Si}{2} line that is undetectable
at our current signal-to-noise ratio.  The strongest optical lines in
\hk\ are the blue \ion{Fe}{3} lines and we do see evidence for
depolarization by 0.2\% and maybe an angle rotation in $\lambda$5129.
The polarization signatures in the blue may be expected to be small
(in the absence of clumping) if all the weak overlapping iron-peak
lines destroy polarization of the pseudo-continuum at blue
wavelengths, so the additional opacity of strong lines has little
additional depolarizing effect.

Although no pre-existing supernova model explains all the
peculiarities of \cx, one new model that does address some of the
issues and makes specific polarimetric predictions for both the
continuum and lines is the ejecta hole model of \citet{ka04}.  They
followed up on the simulations of \citet{mar00}, who determined that 
the donor star to the exploding white dwarf of an SN~Ia could leave a
noticeable imprint on the ejecta in the form of a hole.  The opening
angle of the hole was 30--40\degr\ for main sequence, sub-giant, or
red giant donor stars.  \citet{ka04} found that sightlines down the
hole allow an observer to see more deeply into the ejecta.  This has
several effects on the observed spectrum, including increasing the
prominence of higher-ionization species, weakening absorption lines,
and lowering the observed photospheric velocity relative to more typical
sightlines.  These effects seem promising for SNe 2002cx and 2005hk.
Unfortunately, they also found that objects viewed down a hole
would be somewhat brighter than normal.  Kasen et al. avoided this
difficulty by proposing that if SN 1991T-like objects are
intrinsically normal SNe~Ia viewed down a hole, then \cx -like objects
could be the analogues of the subluminous SN 1991bg-like class of
objects, but viewed down an ejecta hole.  One difficulty with this
scenario is that the SN~2002cx-like objects have only been seen in
late-type spiral galaxies \citep{jha06}, while the SN 1991bg-like
objects are more common in early-type galaxies \citep{how01}.

With that in mind, \citet{ka04} made specific spectropolarimetric
predictions for the consequences of an ejecta hole.  If the system were
oriented such that our line of sight were aimed straight down the
hole, then the projection onto the sky would be symmetrical and very
little polarization should be expected.  Such an exact alignment would
be quite rare and even small angular offsets from the axis of symmetry
should produce measurable effects.  They made a sample
calculation of a polarization spectrum for a line of sight oriented
20\degr\ to the axis of their model hole (opening half-angle of
40\degr ) and found that such an 
orientation gave very low continuum polarization ($\sim$0.1\%,
reflecting the fact that their model was symmetric apart from the
hole) with strong line polarization peaks ($\sim$1\% level).  We
measure a somewhat higher continuum polarization and no strong line
features, in apparent contradiction to the expectations of this model.

\section{Conclusions}

We have presented single-epoch spectropolarimetry of \hk\ and an
analysis of the high signal-to-noise-ratio total-flux spectrum.  The
flux spectrum is very similar to that of the peculiar SN~Ia 2002cx.  Both
supernovae resemble the class of SN 1991T-like objects near maximum light
in that they show prominent lines of \ion{Fe}{3} and only weak lines
of intermediate-mass elements like \ion{Si}{2} and \ion{Ca}{2} that 
characterize the early spectra of normal SNe~Ia \citep{fil97}.  In
addition, \hk\ shows low expansion velocities of $\sim$6000 km
s$^{-1}$, similar to those of \cx.  \citet{li03} found that no
existing theoretical model for SNe~Ia could reproduce all of the
peculiarities of \cx.  Observations of \hk\ provide additional
constraints for models of this new subclass of SNe~Ia.

We have presented a set of line identifications using \synow.  Our
\synow\ fit for \hk\ is very similar to a previous analysis of \cx\
\citep{br04}, but with five additional ionic species.  Four of these
are due to our superior wavelength range and do not necessarily
indicate spectral differences between the two objects.  A detailed
analysis of a time series of spectra could also clarify the
identification of spectral features attributed to those species whose
presence is less certain in our analysis of this individual spectrum.
If all of our possible line identifications are correct, then we have  
identified the signatures of iron-peak elements, intermediate-mass
elements, and maybe unburned carbon at similar velocities in the early
spectra of \hk.  Our \synow-based analysis is unsuitable for
determining the actual relative abundances of these three groups of
species, but a clear next step for future research is for detailed
supernova spectrum synthesis codes to attempt to model the spectrum of
\hk. 

Of particular interest is whether normal deflagration products
(e.g., Nomoto et al. 1984) can fit the \hk\ spectrum for favorable
values of the temperature, density, and ionization. \citet{jha06}
found evidence for the large-scale mixing predicted in
three-dimensional simulations of deflagrations \citep{rhn02,gam03} in 
the presence of low-velocity intermediate-mass elements and possibly
\ion{O}{1} at late times in \cx.  Alternatively, our identification of
\ion{Co}{2} and \ion{Ni}{2} in \hk\ might be evidence for freshly
synthesized $^{56}$Ni and its decay products in the outer layers of
the supernova, as was interpreted in SN 1991T to be the sign of a
detonation \citep{fi92a,maz95}.  The short half-life of
$^{56}$Ni implies that nickel must have been present in larger
quantities at earlier times for it to still be seen in the current
spectrum, so the signatures of nickel should be looked for in any
spectra of \cx-like objects obtained at very early epochs.
A potential consequence is that radioactive material present in the
outermost ejecta might not make a large contribution to the luminosity.

One of the most perplexing mysteries of \cx\ is how that object could
have a spectrum dominated by iron at both early and late times and yet
have such a low luminosity and small expansion velocities.  The
characteristic velocity scale (and hence kinetic energy per unit mass)
of SNe~Ia near maximum light ($>$10,000 km s$^{-1}$) is set by the energy
release of the fusion of carbon and oxygen up to iron-peak or even
just intermediate-mass elements.  The presence of a large fraction of
unburned material that adds inertial mass to the ejecta could solve
this puzzle, but evidence of this raw material in the spectra is
scarce. One argument against a large amount of unburned material is
the rapid spectral evolution seen originally in \cx, and confirmed in
\hk\ by the stronger lines of intermediate-mass elements seen in this
spectrum compared to the one from three days earlier \citep{jha06}.
This may indicate that the outer ejecta have relatively low mass,
though an analysis of the photometry and spectral evolution is
needed to confirm this hypothesis.

One line of evidence in favor of unburned material is the possible
presence of weak \ion{O}{1} lines in the late-time spectra of \cx.
Here we have looked for carbon in the early-time spectrum of \hk\ and
have obtained some potential evidence for \ion{C}{3}.  Features
suggestive of \ion{C}{2} are also present, but we were
unable to find a good fit, though more detailed handling
of non-LTE effects might allow a secure identification to be made.
Similar features seen in normal SNe~Ia have been identified as
\ion{C}{2} in the past \citep{br03,gar04}.
\citet{hof02} were able to show that the peculiar subluminous SN
1999by had unburned material above the photosphere at early times by
detecting \ion{C}{1} lines in the near-IR.  Near-IR spectroscopy of
\hk\ or another member of the class of SN 2002cx-like objects at early
times would be very useful to see if they more closely resemble SN
1999by in this respect or the more normal SN~Ia sample of
\citet{mar06}, which lacked the \ion{C}{1} lines.

Our spectropolarimetric data for \hk\ are perhaps the least unusual
aspect of the object.  They show a low level of continuum polarization
($\sim$0.4\%) and evidence for only weak line features.  Therefore, the
level of asphericity is quite normal for an SN~Ia and models for the
unusual aspects of the \cx-like class of objects that invoke large
deviations from spherical symmetry can be rejected, such as the ejecta
hole model of \citet{ka04}.  The lack of strong line features in the
spectropolarimetry may be related to the weakness of 
the \ion{Si}{2} and \ion{Ca}{2} lines.  If this is correct, SN
1991T-like objects observed in the future may only show weak
polarimetric line features, unless clumping in the ejecta is
important.  Lastly, a full identification of the weak spectral
features might provide additional insight into the nature of \hk.  

\acknowledgments
The authors would like to acknowledge useful conversations with Doug
Leonard.  Comments by Jennifer Hoffman improved the presentation of
this paper.  We also thank the referee, Andy Howell, for a prompt and
useful report.
The data presented herein were obtained at the W.M. Keck Observatory,
which is operated as a scientific partnership among the California
Institute of Technology, the University of California and the National
Aeronautics and Space Administration. The Observatory was made
possible by the generous financial support of the W. M. Keck
Foundation.  The authors wish to recognize and acknowledge the very
significant cultural role and reverence that the summit of Mauna Kea
has always had within the indigenous Hawaiian community.  We are most
fortunate to have the opportunity to conduct observations from this
mountain.  We also would like to thank the expert assistance of the
Keck staff in making these observations possible.  Supernova research
at UC Berkeley is supported by NSF grant AST-0307894.

\clearpage

\begin{deluxetable}{lccccccccccc}
\tablenum{1}
\setlength{\tabcolsep}{4pt}
\tablecaption{\synow\ Fitting Parameters}
\tablehead{\colhead{ } & \colhead{C III} & \colhead{O I} &
\colhead{Mg II} & \colhead{Si II} & \colhead{Si III} & \colhead{S II}
& \colhead{Ca II} & \colhead{Ti II} & \colhead{Fe III} & \colhead{Co
II} & \colhead{Ni II}}
\startdata
$\lambda_{ref}$ (\AA) & 4647 &  7773 & 4481 & 6347 & 4553 & 5454 & 3934 &
4550 & 4420 & 4161 & 4067\\
$\tau$ & 0.3 & 0.08 & 0.35 &  0.2 & 0.3 & 0.25 & 0.8 & 0.04 & 0.7 &
0.2 & 0.1 \\
\label{synowtab}
\enddata
\end{deluxetable}

\clearpage

\clearpage

\clearpage

\clearpage

\clearpage

\end{document}